\documentclass[onecolumn,showpacs,preprintnumbers,amsmath,amssymb]{revtex4}

\usepackage{tabularx}
\usepackage{array}
\usepackage{mathrsfs}
\usepackage{amssymb}
\usepackage{amsmath}
\usepackage{graphicx}
\usepackage{bm}
\usepackage{multirow}
\usepackage{float}
\usepackage{appendix}

\begin{document}
\title{Efficient passive measurement-device-independent quantum key distribution}
\author{Chun-Hui Zhang$^{1,2,3}$}
\author{Chun-Mei Zhang$^{1,2,3}$}
\author{Qin Wang$^{1,2,3}$}\email{qinw@njupt.edu.cn}
\address{ $^1$ Institute of quantum information and technology, Nanjing University of Posts and Telecommunications, Nanjing 210003, China \\
 $^2$ \lq\lq Broadband Wireless Communication and Senser Network Technology\rq\rq Key Lab of Ministry of Education, NUPT, Nanjing 210003, China \\
 $^3$ \lq\lq Telecommunication and Networks\rq\rq National Engineering Research Center, NUPT, Nanjing 210003, China\\}

\date{\today}

\begin{abstract}
The measurement-device-independent quantum key distribution (MDI-QKD) possesses the highest security among all practical quantum key distribution protocols. However, existing multi-intensity decoy-state methods may cause loopholes when modulating light intensities with practical devices. In this paper, we propose a passive decoy-state MDI-QKD protocol based on a novel structure of heralded single-photon sources. It does not need to modulate the light source into different intensities and thus can avoid leaking modulation information of decoy states to eavesdroppers, enhancing the security of practical MDI-QKD. Furthermore, by combining the passive decoy states and biased basis choices, our protocol can exhibit distinct advantages compared with state-of-the art MDI-QKD schemes even when the finite key-size effect is taken into account. Therefore, our present passive decoy-state MDI-QKD seems a promising candidate for practical implementation of quantum key distributions in the near future.
\end{abstract}

\pacs{03.67.Dd, 03.67.Hk, 42.65.Lm}

\maketitle

\section{Introduction}
Quantum key distribution (QKD) \cite{BB84,E91} allows legitimate users, Alice and Bob, to share secret keys under unprecedented level of security based on quantum mechanics. Its unconditional security has been proven with ideal devices \cite{Mayers,Lo1,Shor}. However, the device imperfections in reality offers opportunities for eavesdropper, Eve, to launch quantum hacking \cite{PNS,Blind,Gerhardt}. Fortunately, some countermeasures are invented to fix these security loopholes, such as decoy-state method \cite{Hwang,WXB2005,Lo2} and measurement-device-independent QKD (MDI-QKD) \cite{Braunstein,Lo3}.

In most experimental demonstrations of decoy-state QKD \cite{Qin1,WangS,Boaron421,RFIMDI,Yin404}, people actively modulate light sources into different intensities with acousto- or electro-optic modulators. However, there may exist side-channel loopholes during the intensity modulation processes. For example, when a modulator is not properly designed, some physical parameters of the pulses emitted by the sender may depend on the particular setting selected \cite{Curty1,HuJZ}, causing severe security problems. To reduce the information leakage, some passive decoy-state methods have been put forward \cite{Mauerer,Adachi,Qin2,Curty2,Qin3} and experimentally demonstrated \cite{SunSH,Qin4,Qin5}.

However, in practical implementations of MDI-QKD \cite{WXB1,WXB2,LSA,Curty3,WXB3,WXB4,WXB5,Qin6,WXB6}, only active decoy-state methods are adopted. It is mainly due to the fact that in most passive schemes the vacuum state cannot be directly observed and also hard to precisely estimate with those observed data. To solve the problem, in this paper we propose an improved passive decoy-state MDI-QKD protocol. It is based on a novel structure of heralded single-photon sources (HSPS) presented in our recent work \cite{Qin2,Qin4}, by combining the idea of scaling in \cite{WXB6}, we are able to give more precise parameter estimations with the data that can be observed in experiment. Through carrying out full parameter optimizations, we do investigation on its performance with finite-size-key effects. Corresponding simulation results demonstrate that our present scheme can work more efficiently than state-of-the art decoy-state MDI-QKD scheme using WCS \cite{WXB4}, and can even show some advantages compared with the most advanced MDI-QKD scheme using HSPS \cite{WXB6}.

\section{Passive decoy-state MDI-QKD protocol}
\begin{figure*}[htbp]
\centering
\includegraphics[width=12cm]{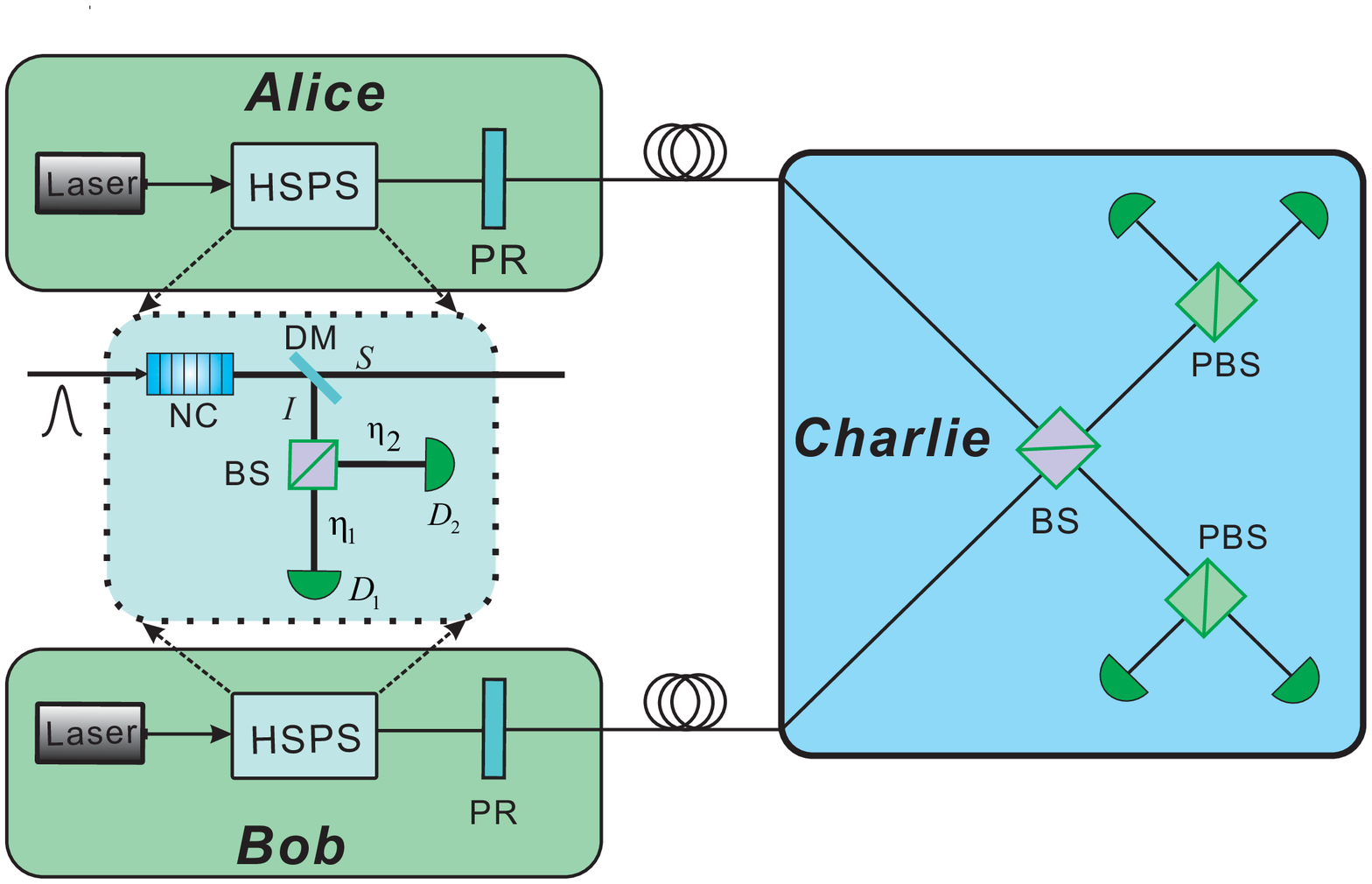}
\caption{The schematic of our passive decoy-state MDI-QKD protocol. HSPS, heralded single-photon source; NC, nonlinear crystal; DM, dichroic mirror; BS, beam splitter; PBS, polarization beam-splitter; PR, polarization rotator; $D_1$ and $D_2$, single-photon detectors.}
\label{Fig1}
\end{figure*}
Our passive decoy-state MDI-QKD protocol is presented in Fig. \ref{Fig1}. At Alice's (Bob's) side, a time correlated pulse pair from parametric down-conversion (PDC) process of nonlinear crystal is split into idler mode $I$ and signal mode $S$. The mode $I$ is further split and then measured by two local detectors, while mode $S$ is encoded conditionally on the local detection events and sent to untrustworthy third party (UTP), Charlie, for bell state measurement (BSM). The local detection events can be divided into four kinds, denoted as $V_i$ $(i = 1, 2, 3, 4)$: (1) Non-clicking; (2) Single clicking at $D_1$; (3) Single clicking at $D_2$; (4) Clicking at both $D_1$ and $D_2$. These events $V_i$ correspond to projected state $l$ $(l = w, x, y, z)$ of mode $S$ in the photon-number space.
In our protocol, when $V_i$ event happens, the signal mode $S$ will be encoded on specific basis, i.e., state $w$ and $x$ are only prepared in $X$ basis while state $y$ and $z$ only prepared in $Z$ basis. The data on $X$ basis is used to estimate channel parameter, and the data on $Z$ basis is used to distill keys. The nomenclature and corresponding relationship is shown in Table \ref{tab1}.

\begin{table}[H]
\centering
\caption{The nomenclature and corresponding relationship in our protocol.}
\begin{tabular} {ccccc}
 \hline \hline
    Events & State & Probability & Encoding basis \\ \hline
    $V_1$ & $w$ & $P_n^w$ & $X$\\
    $V_2$ & $x$ & $P_n^x$ & $X$\\
    $V_3$ & $y$ & $P_n^y$ & $Z$\\
    $V_4$ & $z$ & $P_n^z$ & $Z$\\
 \hline \hline
  \end{tabular}
 \label{tab1}
\end{table}

Condition on event $V_i$, the mode $S$ is projected into the photon-number space ${\rho _l} = \sum\nolimits_n {P_n^l| n \rangle \langle n |}$, and the photon number statistics can be expressed as \cite{Qin2}
\begin{equation}
P_n^l = {P_n}\sum\limits_{{s_1}{s_2}} {{P_{{V_i}|{s_1}{s_2}}}{P_{{s_1}{s_2}|n}}},
\label{Pnl}
\end{equation}
where $P_n$ is the photon-number distribution of PDC process; $P_{s_1s_2|n}$ denotes the projecting probability of $n$-photon state passing through the BS and being projected into state $|s_1s_2\rangle$; $P_{V_i|s_1s_2}$ denotes the probability of an $V_i$ event given a projected state $|s_1s_2\rangle$. Note that $P_n$ can be certain distribution of PDC process, here we assume a Poisson distribution, i.e., ${P_n} = \frac{{{\mu ^n}}}{{n!}}{e^{ - \mu }}$, where $\mu$ is the mean photon number of idler mode or signal mode. Besides, all $P_{V_i|s_1s_2}$ are listed in Table \ref{tab2}, while $P_{s_1s_2|n}$ is given by \cite{Qin2}
\begin{align}
{P_{{s_1}{s_2}|n}} =\sum\limits_{k = 0}^n {\sum\limits_{{s_2} = 0}^{n - k} {\sum\limits_{{s_1} = 0}^k {\frac{{n!{t^k}{{\left( {1 - t} \right)}^{n - k}}\eta _{1}^{{s_1}}\eta _{2}^{{s_2}}{{\left( {1 - {\eta _{1}}} \right)}^{k - {s_1}}{\left( {1 - {\eta _{2}}} \right)^{n - k - {s_2}}}}}}{{{s_1}!{s_2}!\left( {k - {s_1}} \right)!\left( {n - k - {s_2}} \right)!}}} } },
\label{Ps1s2}
\end{align}
where $t$ represents the transmission efficiency of the BS, reasonably assume $t \in (0,{{\textstyle{1 \over 2}}})$; $\eta_{1}$ and $\eta_{2}$ denote the overall efficiency of each branch in the idler mode respectively, which includes the detection efficiency but excludes the transmission efficiency of the BS ($t$); $d_{1}$ and $d_{2}$ refer to the dark count rate of $D_1$ and $D_2$.
\begin{table}[H]
\centering
\caption{Probability of the $V_i$ event occurring.}
\begin{tabular} {cccccc}
 \hline\hline
  Case & $P_{V_1|s_1s_2}$ & $P_{V_2|s_1s_2}$ & $P_{V_3|s_1s_2}$ & $P_{V_4|s_1s_2}$ \\ \hline
    $s_1=0,s_2=0$ & $(1-d_1)(1-d_2)$ & $d_1(1-d_2)$ & $d_2(1-d_1)$ & $d_1d_2$ \\
    $s_1\neq0,s_2=0$ & 0 & $1-d_2$ & 0 & $d_2$ \\
    $s_1=0,s_2\neq0$ & 0 & 0 & $1-d_1$ & $d_1$ \\
    $s_1\neq0,s_2\neq0$ & 0 & 0 & 0 & 1 \\
 \hline \hline
  \end{tabular}
 \label{tab2}
\end{table}

In the MDI-QKD, Alice and Bob simultaneously send photon pulses to the UTP. When Alice sends state $l$ and and Bob sends state $r$ $(l,r \in \{w,x,y,z\})$, we can obtain the gains ${S_{lr}} = \sum\nolimits_{j,k\geqslant0} {a_j^lb_k^r{Y_{jk}}}$, and quantum bit errors ${\kern 1pt} {T_{lr}} = \sum\nolimits_{j,k\geqslant0} {a_j^lb_k^r{Y_{jk}}{e_{jk}}} $. Here $a_j^l$ and $b_k^r$ are the probability distribution as presented in Eq. (\ref{Pnl}) at Alice's side and Bob's side, respectively. $Y_{jk}$ and $e_{jk}$ each denotes the yield and the error rate when Alice sends a $j$-photon state and Bob sends a $k$-photon state.  The average quantum-bit error-rate (QBER) is given by ${E_{lr}} = {{{T_{lr}}} \mathord{\left/{\vphantom {{{T_{lr}}} {{S_{lr}}}}} \right.\kern-\nulldelimiterspace} {{S_{lr}}}}$.
From Ref. \cite{Qin2,WXB1}, we know that former formulae used to estimate the yield and error rate of single-photon pair in $X$ basis are
\begin{align}\label{Y11old}
Y_{11}^{X,L} = \frac{{a_1^xb_2^x\left( {\underline{S}_{ww} - \overline{{\cal H}}} \right) - a_1^wb_2^w\left( {\overline{S}_{xx} - \underline{{\cal H}}'} \right)}}{{a_1^wa_1^x(b_1^wb_2^x - b_1^xb_2^w)}}, \quad
e_{11}^{X,U} = \frac{{\overline{T}_{ww} - {\textstyle{1 \over 2}}\underline{{\cal H}}}}{{a_1^wb_1^wY_{11}^{X,L}}},
\end{align}
where ${\cal H} \equiv a_0^wb_0^w{Y_{00}} + \sum\nolimits_{m = 1}^\infty  {(a_0^wb_m^w{Y_{0m}} + a_m^wb_0^w{Y_{m0}})}$, ${\cal H}' \equiv a_0^xb_0^x{Y_{00}} + \sum\nolimits_{m = 1}^\infty  {(a_0^xb_m^x{Y_{0m}} + a_m^xb_0^x{Y_{m0}})} $; $L$ and $U$ refer to the lower bound and upper bound, respectively. Besides, we denote overline and underline as the experimental values with statistical fluctuation. For example, for experimental value of ${S_{ww}}$, it satisfies ${\underline{S}_{ww}} := {S_{ww}} - {\Delta _1} \leqslant {\widetilde{S}_{ww}}  \leqslant {\overline{S}_{xx}} := {S_{xx}} + {\Delta _2}$ with a failure probability $\varepsilon$, where $\Delta _1$ and $\Delta _2$ are the statistical fluctuation values.

However, in fact, the experimental values related to vacuum state presented in Eq. (\ref{Y11old}), i.e., ${\cal H}$ and ${\cal H}'$, cannot be directly measured in our passive MDI-QKD protocol. Fortunately, we can borrow the idea in \cite{WXB6} to reformulate Eq. (\ref{Y11old}) as
\begin{align}\label{Y11new}
\widetilde{Y}_{11}^{X,L}(\widetilde{{\cal H}}) = \frac{{a_1^xb_2^x {\underline{S}_{ww}}  - a_1^wb_2^w{\overline{S}_{xx}} - a_1^xb_2^x \widetilde{{\cal H}} }}{{a_1^wa_1^x(b_1^wb_2^x - b_1^xb_2^w)}},
\end{align}
and
\begin{align}\label{e11new}
\widetilde{e}_{11}^{X,U}(\widetilde{{\cal H}}) = \frac{{\overline{T}_{ww} - {\textstyle{1 \over 2}}\widetilde{{\cal H}}}}{{a_1^wb_1^w\widetilde{Y}_{11}^{X,L}(\widetilde{{\cal H}})}}.
\end{align}
where $\widetilde{{\cal H}}$ is the value of ${\cal H}$, which is a joint parameter existing in $Y_{11}$ and $e_{11}$. According to the $\widetilde{e}_{11}^{X,U}(\widetilde{{\cal H}})$ in Eq. (\ref{e11new}), we can obviously have the range of $\widetilde{{\cal H}}$ as
\begin{align}\label{Hrange}
\widetilde{{\cal H}} \in \left[ {0,2{\overline{T}_{ww}}} \right].
\end{align}
In our protocol, we use the data in $X$ basis to estimate the channel parameters in $Z$ basis ($Y_{11}^Z$ and $e_{11}^{ph}$), but their real values in $Z$ basis may deviate from the quantities in $X$ basis. Therefore, a stricter treatment should be taken here. We use
\begin{align}\label{Ye11ZX}
 - {\Delta _1^\prime} \leqslant Y_{11}^{Z}(\widetilde{{\cal H}}) - \widetilde{Y}_{11}^{X,L}(\widetilde{{\cal H}}) \leqslant {\Delta _2^\prime}, \\
 - {\Delta _1^{\prime\prime}} \leqslant e_{11}^{ph}(\widetilde{{\cal H}}) - \widetilde{e}_{11}^{X,U}(\widetilde{{\cal H}}) \leqslant {\Delta _2^{\prime\prime}},
\end{align}
for the worst-case values of $Y_{11}^{Z,L}(\widetilde{{\cal H}})$ and $e_{11}^{ph,U}(\widetilde{{\cal H}})$ with a failure probability.

With the above formulae, we shall estimate the worst-case result for the key rate by scanning all possible values of ${\widetilde{{\cal H}}}$ given in Eq. (\ref{Hrange}), and get the final key generation rate per pulse as
\begin{align}\label{SKR}
R\geqslant\mathop {min}\limits_{\widetilde{{\cal H}}} R(\widetilde{{\cal H}}) = \left\{ {a_1^yb_1^yY_{11}^{Z,L}(\widetilde{{\cal H}})\left[ {1 - {H_2}\left( {e_{11}^{ph,U}(\widetilde{{\cal H}})} \right)} \right] - S_{yy}^Zf{H_2}\left( {E_{yy}^Z} \right)} \right\} - \frac{1}{{{N_t}}}\left( {{{\log }_2}\frac{8}{{{\epsilon _{cor}}}} + 2{{\log }_2}\frac{2}{{\varepsilon '\hat \varepsilon }} + 2{{\log }_2}\frac{1}{{{\varepsilon _{PA}}}}} \right),
\end{align}
where $f$ is the inefficiency of error correction; $H_2(p)=-p\log_2(p)-(1-p)\log_2(1-p)$ is the binary Shannon entropy function. The second part of Eq. (\ref{SKR}) is due to other finite size effects on the key rate as defined in \cite{Curty3}, guaranteeing the composable security of whole MDI-QKD system. Note that here we only use the state $y$ conditional on event $V_3$ to generate keys. Though we can also use state $z$ to distill keys, our numerical simulation shows that it raise the key rate little because the probability of event $V_4$ is very low compared with event $V_3$.

\section{Simulation RESULTS}
In the following, we carry out numerical simulations for our proposed passive decoy-state MDI-QKD protocol. Here, we focus on the symmetric case, which means that Charlie is at the middle of Alice and Bob, and all other device parameters of Alice's side and Bob's side are identical. Let $d_1 = d_2 = d_A$ and $\eta_1 = \eta_2 = \eta_A$, we can obtain the simplified photon-number distribution for $a_n^l, b_n^r$ ($l,r\in\{w,x,y,z\}$) as
\begin{align}\label{Pwxyz}
\begin{array}{l}
P_n^w = {(1 - {d_A})^2}{(1 - {\eta _A})^n}\frac{{{\mu ^n}}}{{n!}}{e^{ - \mu }},\\
P_n^x = (1 - {d_A}){(1 - {\eta _A})^n}\left[ {{{\left( {\frac{{1 - (1 - t){\eta _A}}}{{1 - {\eta _A}}}} \right)}^n} + {d_A} - 1} \right]\frac{{{\mu ^n}}}{{n!}}{e^{ - \mu }},\\
P_n^y = (1 - {d_A}){(1 - {\eta _A})^n}\left[ {{{\left( {\frac{{1 - t{\eta _A}}}{{1 - {\eta _A}}}} \right)}^n} + {d_A} - 1} \right]\frac{{{\mu ^n}}}{{n!}}{e^{ - \mu }},\\
P_n^z = \frac{{{\mu ^n}}}{{n!}}{e^{ - \mu }} - P_n^w - P_n^x - P_n^y.
\end{array}
\end{align}
Here, the Eq. (\ref{Pwxyz}) needs to satisfy the condition ${\textstyle{{{P_n^x}} \over {{P_n^w}}}} \geqslant {\textstyle{{{P_2^x}} \over {{P_2^w}}}} \geqslant {\textstyle{{{P_1^x}} \over {{P_1^w}}}}$ for $n\geqslant3$ to let Eq. (\ref{Y11new}) hold, and the proof of this condition has been given in \cite{Qin2}.

\begin{figure}[htbp]
\centering
\includegraphics[width=11cm]{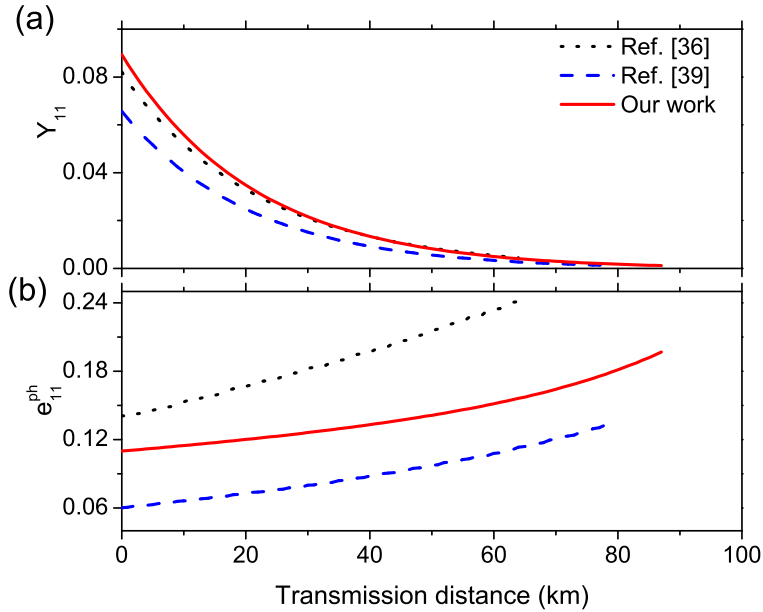}
\caption{The single-photon-pair yield ($Y_{11}$) and phase-flip error rate ($e_{11}^{ph}$) of different schemes. Here the data size $N_t=10^9$.}
\label{Fig2}
\end{figure}

\begin{table}[htbp]
  \caption{The device parameters used in our numerical simulations. $e_0$, the error rate of vacuum pulses; $e_d$, the misalignment error; $\eta_A$ and $d_A$, the overall efficiency and the dark count rate of the local detector at Alice (or Bob)'s side; $\eta_C$ and $d_C$ each corresponds to the detection efficiency and the dark count rate of detectors at the Charlie's side; $f$, the inefficiency of error correction.}
\renewcommand{\arraystretch}{1.3}
\begin{tabularx}{\linewidth}{XXXXXXXX}  \hline \hline
$e_0$ &  $e_d$ &   $\eta_A$ &  $d_A$   & $\eta_C$ &  $d_C$ & $f$\\ \hline
0.5 &  1.5\% &  75\% & $10^{-6}$  &  40\% &  $10^{-7}$ & 1.16\\
    \hline\hline
  \end{tabularx}
 \label{tab3}
\end{table}
In our simulation, we use the commonly used loss coefficient of 0.2 dB/km, and other device parameters are listed in Table \ref{tab3}. We make comparisons between our present work and other two representative active MDI-QKD schemes \cite{WXB4,WXB6}, in which Ref. \cite{WXB6} is the best active MDI-QKD scheme so far. Furthermore, due to the finite-size key effect, we adopt the same statistical fluctuation method (normal distribution given the failure probability $\varepsilon=10^{-7}$) and composable security level as in \cite{WXB6}. Besides, we perform full parameter optimization for all schemes in comparisons. In our protocol, it includes mean photon number $\mu$, transmission efficiency $t$ of the local BS, and the joint parameter $\widetilde{{\cal H}}$. The comparisons are shown in Figs. \ref{Fig2} $\sim$ \ref{Fig4}.

\begin{figure}[h!]
\centering
\includegraphics[width=12cm]{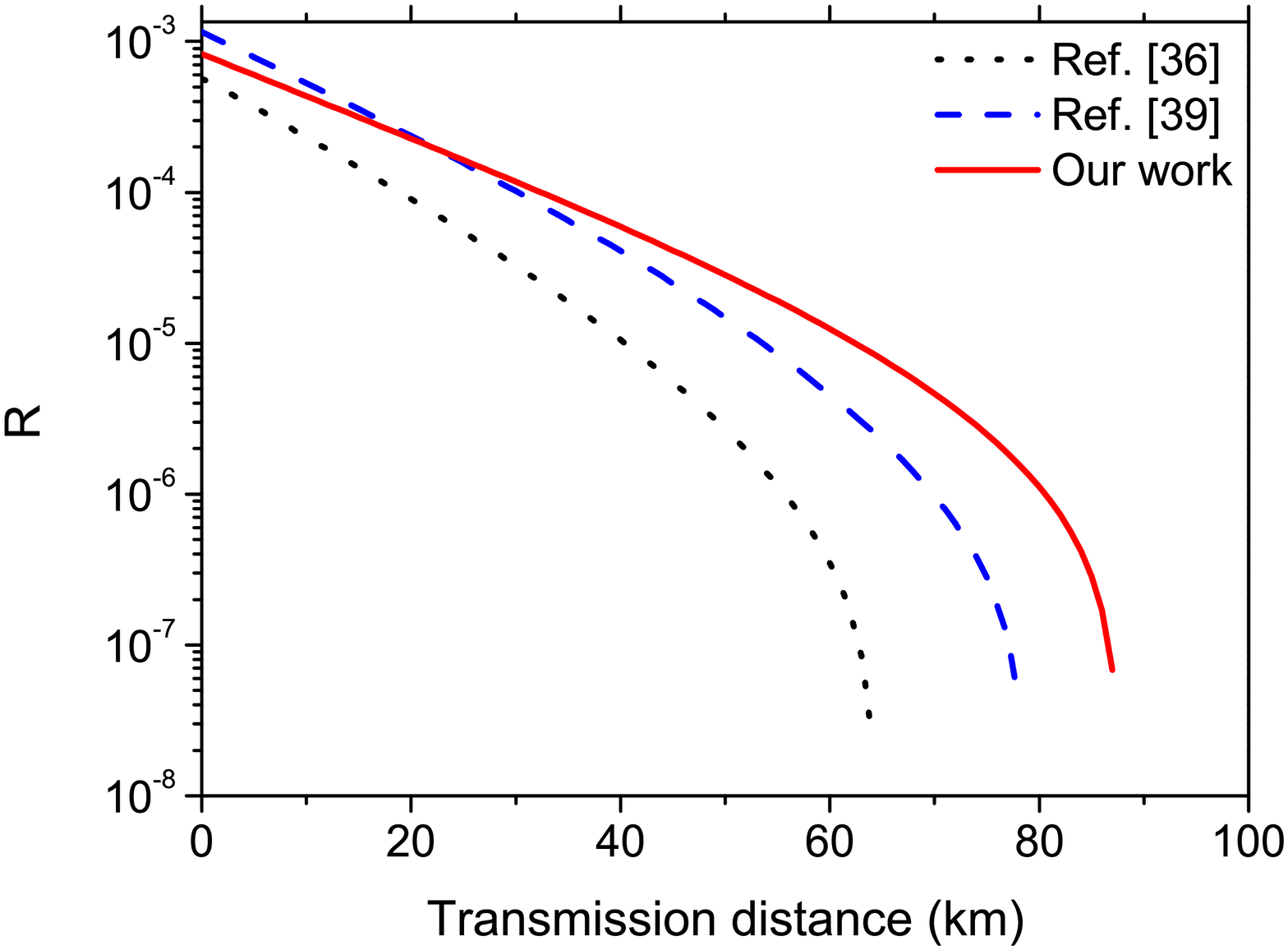}
\caption{The key rates of our scheme and other two reports at data size $N_t=10^9$.}
\label{Fig3}
\end{figure}

\begin{figure}[htbp]
\centering
\includegraphics[width=12cm]{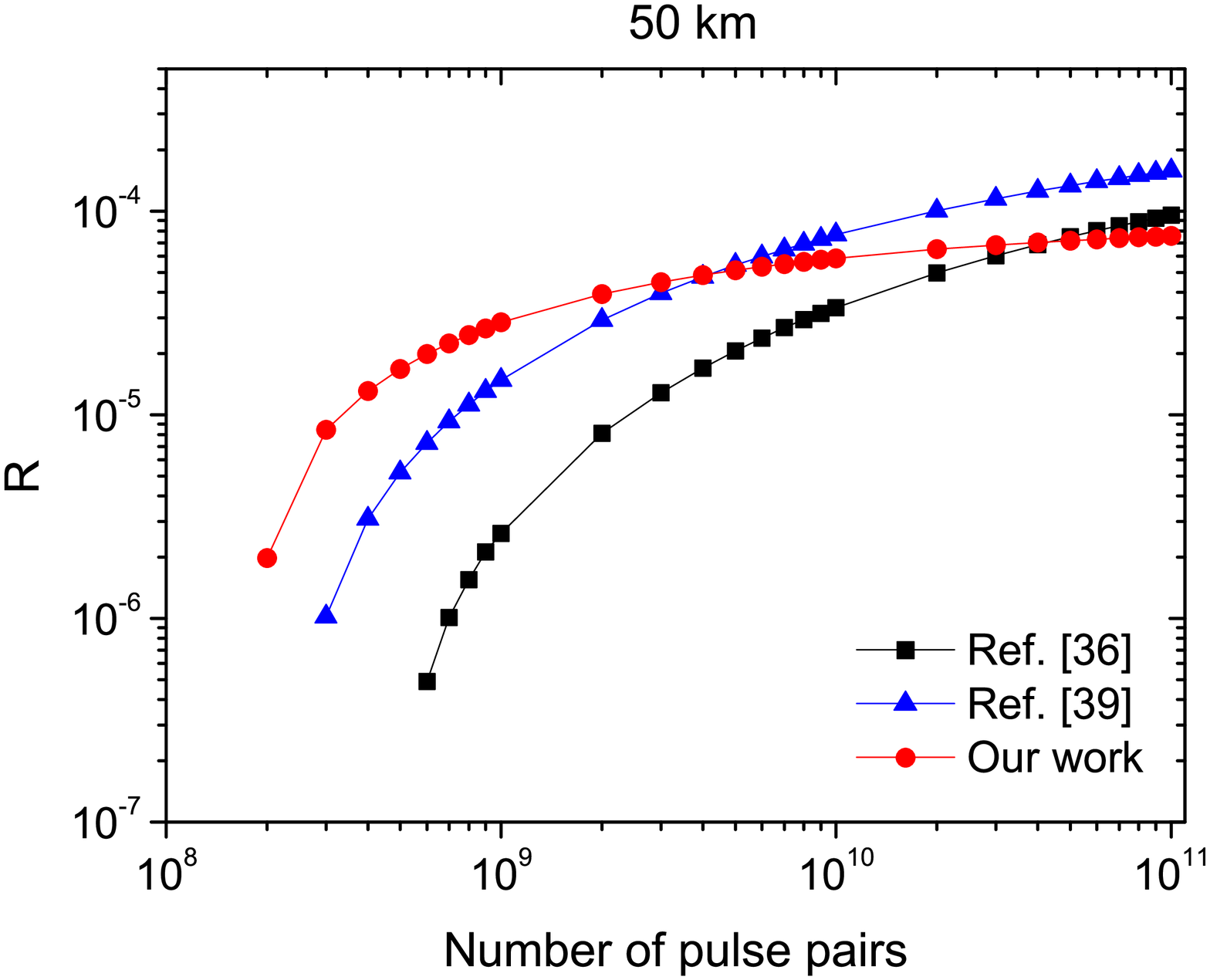}
\caption{The key rates of different schemes versus data size at 50 km.}
\label{Fig4}
\end{figure}

We first plot the comparisons for two key channel parameters, i.e., the yield and the phase-flip error rate of single-photon pairs, between our protocol and other two schemes in Fig. \ref{Fig2}. Here the data size of the pulse number each side sends is reasonably set as $N_t=10^9$. Fig. \ref{Fig2} (a) shows that $Y_{11}$ is estimated more accurately in our passive protocol than other two active schemes, which mainly attributes to that the local detection structure provides some photon-number-resolution ability, while Fig. \ref{Fig2} (b) shows the estimated $e_{11}^{ph}$ of our protocol is in the middle of other two schemes. Finally, the comparisons of key rates for different schemes are illustrated in Fig. \ref{Fig3}. It shows that our protocol always outperforms Ref. \cite{WXB4} at this data size, and will be advantageous at a relatively longer transmission distance compared with Ref. \cite{WXB6}, i.e., $>30$ km.

Besides, we also investigate the key generation rate changing with the variation of the data size at fixed transmission distance (50 km), as shown in Fig. \ref{Fig4}. From Fig. \ref{Fig4}, we find that our present work has cross points with the other two state-of-the art MDI-QKD works using either WCS and HSPS, and show best performance at small data size, e.g., $N_t\leqslant4\times10^{9}$. Compared with Ref. \cite{WXB4}, the improvement mainly due to the inherent merit of the sources as explained in Ref.  \cite{Qin7}. However, even when using the same light sources, HSPS, our work can still exhibit less sensitivity than Ref. \cite{WXB6}, making it very promising candidate for practical applications.

\section{Conclusions and Discussions}
In this paper, we propose a passive decoy-state MDI-QKD protocol, which is impossible for Eve to distinguish between decoy and signal states, enhancing the security of MDI-QKD. The passive decoy states conditional on local events combining the biased basis choice and the idea of scaling in \cite{WXB6} are adopted to estimate channel parameters more accurately. Furthermore, we perform full parameter optimization for our protocol, and simulation results show that our protocol can show some advantages even compared with some state-of-the art MDI-QKD schemes, especially at a longer distance or a smaller data size. Hence, our present work provides an alternative and useful approach to improve both the security and practicability of MDI-QKD protocols, and represents a further step along practical implementations of quantum key distributions.

\section*{ACKNOWLEDGMENTS}
 We gratefully acknowledge the financial support from the National Key R\&D Program of China through Grant Nos. 2017YFA0304100, 2018YFA0306400, the National Natural Science Foundation of China  through Grants Nos.  61475197, 61590932, 11774180, 61705110, 11847215, the Natural Science Foundation of the Jiangsu Higher Education Institutions through Grant No. 17KJB140016, the Natural  Science Foundation of Jiangsu Province through Grant No. BK20170902, the Priority Academic Program Development of Jiangsu Higher Education Institutions, and the Postgraduate Research and  Practice Innovation Program of Jiangsu Province through Grant No. KYCX17\_0792.

\end{document}